\documentstyle[preprint,aps]{revtex}

\preprint{\tt cond-mat/9404026}
\begin{document}
\draft 

\title{Crossover Driven by Time-reversal Symmetry Breaking in Quantum
  Chaos}


\author{N. Taniguchi\cite{address-j}, A.
  Hashimoto~\cite{aki-address}, B. D. Simons and B. L. Altshuler}
\address{Department of Physics, Massachusetts Institute of Technology,
  77 Massachusetts Avenue, Cambridge, Massachusetts 02139 }

\date{11 April 1994} 

\maketitle

\begin{abstract}
  Parametric correlations of energy spectra of quantum chaotic systems are
  presented in the orthogonal-unitary and symplectic-unitary crossover
  region.  The spectra are allowed to disperse as a function of two
  external perturbations: one of which preserves time-reversal symmetry,
  while the other violates it.  Exact analytical expressions for the
  parametric two-point autocorrelation function of the density of states
  are derived in the crossover region by means of the supermatrix method.
  For the orthogonal-unitary crossover, the velocity distributions is
  determined and shown to deviate from Gaussian.

\end{abstract}


\narrowtext 
Spectral properties of quantum chaotic systems are well described by
the Wigner-Dyson statistics of random matrix
theory~\cite{MehtaHaake}.  Originally introduced to study
complex nuclei, random matrix theory (RMT) was found to be relevant to the
spectral statistics of disordered metallic particles (quantum
dots)~\cite{Gorkov65,Efetov83,Altshuler86b}, as well as quantum chaotic
systems in general (e.g., chaotic billiards)~\cite{Bohigas91}.  The most
striking feature is universality: when energies scaled by the
mean level spacing, their distribution depends only on the
symmetry of the Hamiltonian, irrespective of microscopic details.
Typically systems belong to one of three universality classes:
orthogonal (spinless) and symplectic (with spin-orbit interaction) for
those which are invariant under time-reversal invariant systems, and
unitary for those which are not.  
Starting from the Schr\"odinger equation with a random potential, Efetov
has developed a field theoretic description of disordered metals, based on
a supermatrix nonlinear $\sigma$-model~\cite{Efetov83}.  He demonstrated
that, up to the Thouless energy, the density of states correlators for the
three
universality classes are identical to those derived from RMT.  

Subsequently Pandey and Mehta extended RMT to examine crossover behavior
where the symmetry of the system gradually changes from one pure symmetry
to another~\cite{PandeyMehta}.
They obtained an analytical expression for the density of states
correlator, i.e., the two-point cluster function, and related quantities.
(A result recently rederived recently by the supermatrix
method~\cite{Altland93}.) 
The resulting description of the orthogonal-unitary crossover was
numerically justified for a disordered metallic ring with Aharonov-Bohm
flux~\cite{Dupuis91}.

A second approach to quantum chaos is to examine correlations of a
Hamiltonian $H(X)$ which depend on an external parameter $X$.  Beginning
with the work of Pechukas\cite{Pechukas83} and Yukawa~\cite{Yukawa85},
problems of this kind have attracted great interest most
recently~\cite{Wilkinson,Parametric}.  By making use of an appropriate
rescaling, parametric correlations of the density of states as well as
response functions were also shown to be
universal~\cite{Simons,Taniguchi}.

In this Letter, we will examine the universal parametric correlations in
the orthogonal-unitary crossover region.  For
completeness, we mention (but do not discuss) the result for
symplectic-unitary crossover.  We are motivated by problems where the
Hamiltonian $H(X_o,X_u)$ depends on two external parameters $X_o$ and
$X_u$, where the former preserves the
time-reversal symmetry whereas the second violates it.  
Although the spectral statistics of $H(X_o,X_u=0)$ belong to the
orthogonal ensemble, gradually driven to unitary by introducing
T-invariance breaking parameter, $X_u$. 

Parametric correlations in the orthogonal-unitary crossover are relevant to
many problems.  
In a recent experiment~\cite{Sivan94}, the correlator of differential
conductance was measured as a function of magnetic field, $X_u$, in a
heavily doped quantum dot.  These measurements were directly related to
density correlators in the crossover region, considered below.  At the
same time, one can imagine gate voltage serving as second parameter $X_o$,
which perturbs the system but conserves T-invariance.  A second example
could involve an irregular ballistic cavity, or billiard in which a
potential, $X_o$, changes the shapes of the boundary, while a magnetic
field acts as a T-breaking perturbation.  

Our aim here is to determine the parametric autocorrelator of density of
states
\begin{equation}
K(\Omega ,X_o,X_u;\bar X)=\left\langle {\nu (E,\bar X_o-{\textstyle{{X_o}
\over 2}},\bar X-{\textstyle{{X_u} \over 2}})\,\nu (E+\Omega ,\bar
X_o+{\textstyle{{X_o} \over 2}},\bar X+{\textstyle{{X_u} \over 2}})}
\right\rangle,
\end{equation}
where $\nu (E,X_o,X_u)=\sum_n {\delta (E-E_n(X_o,X_u))}$ is the density of
states and $E_n(X_o,X_u)$ are eigenenergies of the Hamiltonian
$H(X_o,X_u)$.  The statistical average denoted by $\langle\cdots\rangle$
can be performed over certain interval of the energy and/or external
parameters.  $K$ depends on the average of the unitary parameter $\bar X$,
as well as on differences of parameters $X_{o,u}$ and energies, $\Omega$.
$\bar X$ serves as a crossover parameter from orthogonal to unitary
symmetry.

Following Ref.~\cite{Simons,Taniguchi}, we switch from $\{\Omega, X_o,
X_u, \bar X\}$ to dimensionless variables $\{\omega, x_o, x_u,\bar x\}$
(conventionally written as lower case),
\begin{equation}
\omega =\Omega / \Delta ;\quad\varepsilon_n=E_n/\Delta
\label{def:omega}
\end{equation}
\begin{equation}
x_o=\sqrt {C_o(0;\bar{X}=0)}\,X_o;\quad x_u=\sqrt
{C_u(0;\bar{X}\rightarrow\infty)}\,X_u;\quad \bar{x}=\sqrt
{C_u(0;\bar{X}\rightarrow\infty)}\,\bar{X},
\label{def:x}
\end{equation}
\begin{equation}
  C_{o,u}(0,\bar X)=\left\langle {\left(
    {{{\partial \varepsilon_n(X_o,X_u)} \over \partial X_{o,u}}} \right)^2}
\right\rangle,
\label{def:C0}
\end{equation}
where $\Delta=\langle\nu\rangle^{-1}$ is the mean level spacing of the
spectrum.
We note that rescaling for $X_o$ is defined in the orthogonal limit,
and for $X_u$ and $\bar X$ at the unitary limit, which follows the
conventions of both.  
We show that a dimensionless correlator defined by
\begin{equation}
  k(\omega ,x_o,x_u;\bar x) = \Delta^2 K(\Omega ,X_o,X_u;\bar X)-1,
\end{equation} 
is a universal function of all dimensionless variables $\omega$,
$x_{o,u}$, and $\bar x$.

Consider the $N\times N$ random matrix $H(X_o,X_u) = H_0+X_o H_s+X_u H_a$,
where $H_0$ and $H_s$ are real symmetric matrices and $H_a$ is an
antihermitian matrix~\cite{PandeyMehta}, the transition from orthogonal to
unitary symmetry for $K(\Omega ,X_o,X_u;\bar X)$ occurs at $\bar X$ of the
order of $1/\sqrt N$ or equivalently $\bar x\sim 1$.  Thus all
non-universal effects such as $\bar X$-dependence of $\Delta$ are
negligible in the limit $N\rightarrow\infty$.  As a result, crossover
behavior of $k(\omega ,x_o,x_u;\bar x )$ is universal.

To evaluate the $k(\omega ,x_o,x_u;\bar x )$, we use the zero dimensional
supermatrix nonlinear $\sigma$-model, which can be derived from the
disordered metallic grain with large dimensionless conductance.  This
model serves as an underlying {\em universal} model for describing
spectral statistics of quantum chaotic systems~\cite{Simons,Taniguchi}
even in the orthogonal-unitary crossover region~\cite{Altland93}.
Following the notation of Ref.~\cite{Efetov83}, $k(\omega ,x_o,x_u;\bar
x)$ can be expressed through the nonlinear $\sigma$-model as 
\begin{eqnarray}
&&k(\omega ,x_u,x_o;\bar x)=-\mbox{Re}\, \int{\cal D}Q\, e^{-F[Q]}{(Q-\Lambda
)_{33}^{11}(Q-\Lambda )_{33}^{22}},\\
&&F[Q]={{i\pi \omega } \over 4}\mbox{STr}\left( {\Lambda Q} \right)-{{\pi
^2} \over 8}\mbox{STr}\left[ {Q,\bar x\tau _3+{{x_u} \over 2}\tau _3\Lambda }
\right]^2-{{\pi ^2x_o^2} \over {64}}\mbox{STr}\left[ {\Lambda ,Q} \right]^2,
\end{eqnarray}
where $Q$ is a supermatrix.  The extension to parametric correlation
functions is straightforwardly done~\cite{supermatrix}.  Performing the
definite integral over the supermatrix, we obtain for orthogonal-unitary
crossover: 
\begin{equation}
k_{ou}(\omega ,x_o,x_u;\bar x)=\mbox{Re}\int_{-1}^1\!d\lambda
\int_1^\infty\!d\lambda_1\int_1^\infty\!d\lambda_2\;e^{-F}\;W_{+}(\omega
,x_o,x_u;\bar x),
\label{k-result}
\end{equation}
where 
\begin{eqnarray}
F=&&-i\pi \omega |\lambda _1\lambda _2-\lambda |+\pi ^2\bar x^2|2\lambda
_2^2-\lambda ^2-1|+{{\pi ^2x_u^2} \over 4}|2\lambda _1^2-\lambda
^2-1|\nonumber\\ &&+{\pi ^2x_{o}^2 \over 4}|2\lambda _1^2\lambda
_2^2-\lambda ^2-\lambda _1^2-\lambda _2^2+1|,
\end{eqnarray}
\begin{eqnarray}
  &&W_{\pm}(\omega ,x_o,x_u;\bar x)= {(\lambda_1\lambda_2-\lambda)^2 \over
    (\lambda _1^2+\lambda _2^2+\lambda ^2-2\lambda \lambda _1\lambda
    _2-1)} \left[\pm {(1-\lambda^2)\cosh \alpha -(\lambda _1^2-\lambda
    _2^2)\sinh\alpha \over (\lambda _1^2+\lambda _2^2+\lambda
    ^2-2\lambda \lambda _1\lambda _2-1) }\right.\nonumber\\ 
&&\qquad \left.
  +\pi^2\left[ (4\bar x^2\lambda _2^2-x_u^2\lambda _1^2)\sinh \alpha
    +(2\bar x^2e^{-\alpha}+{x_u^2\over 2}e^\alpha)(1-\lambda^2)
\right] \right\},
\end{eqnarray}
and we define $\alpha = \pi^2
(\bar x^2-x_u^2/ 4) |1-\lambda ^2|$. 
For completeness, we include the analogous expression for
symplectic-unitary crossover:
\begin{equation}
k_{su}(\omega ,x_o,x_u;\bar x)=\mbox{Re}\int_1^\infty\!d\lambda
\int_{-1}^1\!d\lambda_1\int_0^1\!d\lambda_2\;e^{-F}\;W_{-}(\omega
,x_o,x_u;\bar x),
\end{equation}
where $x_o$ plays a role as any T-invariance perturbation parameter in
this case.  
Hereafter we restrict ourselves to the orthogonal-unitary crossover case
for brevity.

We first check various limiting behaviors of Eq.~(\ref{k-result}).  The
orthogonal limit straightforwardly gives the same result of
Ref.~\cite{Simons} when we set $\bar x, x_u\to 0$.  In the unitary limit
($\bar x\rightarrow\infty$, and $x_o\rightarrow 0$), we can recognize the
leading contribution coming from the term proportional to $\bar x^2\exp
\left( {-F+\alpha } \right)$.  After $\lambda _2$ integration, the
expression is reduced to the same expression for the pure unitary case in
Ref.~\cite{Simons}.  Another interesting limit is the large $\omega$
asymptotics, which can be obtained by expanding the integrand of
Eq.~(\ref{k-result}) around $\lambda =\lambda _1=\lambda _2=1$, and
replacing the integral region over $\lambda $ by $(-\infty ,1)$.  After
straightforward but lengthy calculation, we can show that the large
$\omega $-asymptotics acquires the diffusive form, which can be
interpreted as the Diffuson and Cooperon modes in disordered systems:
\begin{equation}
  k(\omega ,x_u,x_o;\bar x)\cong{1 \over 2\pi^2}\mbox{Re}\left[\left(
  {-i\omega +\pi x_o^2/2+\pi x_u^2} \right)^{-2}+\left( {-i\omega
    +\pi x_o^2/2+4\pi\bar x^2} \right)^{-2}\right].
\end{equation}

We can use $k(\omega ,x_u,x_o;\bar x)$ to obtain the velocity (i.e.,
single level current $\partial \varepsilon_n/\partial x_{o,u}$)
distribution functions.     
Here we can define the two kinds of rescaled velocities both in the
orthogonal and unitary directions by ${\partial\varepsilon_n}/{\partial
  x_{o,u}}$.  Their probability distribution can be obtained from
the formula 
$f_{o,u}(v;\bar x) \equiv \left\langle{\delta\left(
  v-\partial\varepsilon_n/\partial x_{o,u}\right)}
\right\rangle = \lim_{x_o,x_u\to 0}x_{o,u}\,k(\omega =vx_{o,u},\:
x_o,x_u;\bar{x})$~\cite{Kravtsov92}.  
Substituting our exact result Eq.  (\ref{k-result}), we obtain 
\begin{mathletters}
\begin{eqnarray}
  &&f_o(v,\bar x)={1 \over \sqrt{2\pi}}\int_0^\infty {dt\,\left[
    {1+{{1-A(\bar x)} \over {2\pi ^2\bar x^2}}\left( {t-1} \right)}
  \right]\,\sqrt {{{t+2\pi ^2\bar x^2} \over {t+\pi ^2\bar x^2}}}\;\exp
  \left[ {-t-{{v^2} \over 2}\left( {{{t+2\pi ^2\bar x^2} \over {t+\pi
          ^2\bar x^2}}} \right)} \right]},\\ &&f_u(v,\bar x)={e^{-v^2/2}
  \over \sqrt{2\pi}}\left[ {1-A(\bar x) +2\pi^2\bar x^2 A(\bar
  x)\int^\infty_1{dt \sqrt{t}\;e^{(v^2/2+2\pi^2\bar x^2)(1-t)}} }\right],
\end{eqnarray}  
\end{mathletters}
where $A(\bar x)\equiv e^{-2\pi^2\bar x^2}\int_0^1 {e^{2\pi^2\bar
x^2\lambda^2}\,d\lambda }$. The function $A(\bar x)$ behaves like
$1-4\pi^2\bar x^2/3$ for small $\bar x$, but vanishes exponentially for
large $\bar x$.  
Their limiting behaviors are easy to understand.  In the unitary limit
($\bar x\to\infty$), the velocity distribution has a Gaussian in both
directions, 
\begin{equation}
f_o(v;\bar x\to \infty )={\exp \left[ {-v^2}\right]\over\sqrt{\pi}}
\:;\quad
f_u(v;\bar x\to \infty )={\exp \left[ {-v^2/ 2}\right]\over\sqrt
{2\pi }}.
\end{equation}
This means that there is no qualitative difference between the unitary and
the orthogonal perturbations, since T-invariance is already fully broken
in the unitary limit.  On the other hand, in the orthogonal limit, the two
velocity distributions behave quite differently.
\begin{equation}
  f_o(v;\bar x\to 0)={{\exp \left[ {-v^2/2} \right]} \over \sqrt{2\pi}}
\:;\quad 
  f_u(v;\bar x\to 0)=\delta (v).
\end{equation}
$f_u(v;\bar x\to 0)$ collapses to a $\delta$-function, since
$\varepsilon(x_o,x_u) - \varepsilon(x_o,0) \propto x_u^2$ for
$x_u\rightarrow 0$.  However there is nothing special in the orthogonal
direction, so that its distribution is again Gaussian with twice as wide
variance as the unitary limit.  
The velocity distribution interpolates smoothly between these two limiting
cases.  

Since the average velocities vanish, the velocity distributions in the
crossover region can be characterized by their variances: $c_{o,u}(0;\bar
x)\equiv\left\langle(\partial\varepsilon_n/\partial x_{o,u})^2
\right\rangle$.  According to Eq.~(11), they are
\begin{mathletters}
\begin{eqnarray}
  c_o(0;\bar x)&&={{1+A(\bar x)} \over 2}+{{(2\pi ^2\bar x^2+1)A(\bar
      x)-1} \over 2}\mbox{Ei}(-2\pi ^2\bar x^2)\exp (2\pi ^2\bar x^2), 
\label{co-result}\\
  c_u(0;\bar x)&&=1-\left[ {1+2\pi ^2\bar x^2\exp (2\pi ^2\bar
    x^2)\,\mbox{Ei}(-2\pi ^2\bar x^2)} \right]\,A(\bar x), 
\label{cu-result} 
\end{eqnarray}
\end{mathletters}
\noindent where $\mbox{Ei}(-z)=\int^\infty_z dt e^{-t}/t$ is the
exponential integral function.  Hence in the limiting case for $\bar x\ll
1$, we find
\begin{mathletters}
\begin{eqnarray}
  c_o(0;\bar x\to +0)&&\cong 1+{{\pi ^2\bar x^2} \over 3}\ln (2\pi
  ^2\bar x^2),\\
  c_u(0;\bar x\to +0)&&\cong -2\pi ^2\bar x^2\ln (2\pi ^2\bar x^2).
\end{eqnarray}  
\end{mathletters}
\noindent These logarithmic dependencies can be understood by a $2\times 2$
random matrix model~\cite{Kamenev93}, since the nearest neighboring
pairs of energy levels make the dominant contribution for small $\bar x$.  
Around the orthogonal limit, eigenenergies can be considered as
$\varepsilon (\bar x)\approx\sqrt{\varepsilon^2(0)+\bar x^2}$, where
$\varepsilon$ is assumed to obey the Gaussian orthogonal ensemble.
Therefore $c_u(0;\bar x)$ can be evaluated as
\begin{equation}
c_u(0;\bar x)\approx \int_0^\infty {d\varepsilon\,\left( {\partial
\varepsilon/ \partial \bar x} \right)^2P_{\rm GOE}(\varepsilon)}\approx
\int_{\bar x}{d\varepsilon\,\bar x^2/ \varepsilon}\approx -\bar
x^2\ln (\bar x).
\end{equation}
This demonstrates that the logarithmic dependence is specific to the
orthogonal-unitary crossover.  By contrast, $c_u(0;\bar x)\approx
\bar{x}^5$ in the symplectic-unitary crossover because $P_{\rm
  GSE}(\varepsilon ) \sim\varepsilon^4$ for small $\varepsilon$.

\begin{figure}
\vspace{2cm}
\input{epsf.tex}
\centerline{\epsfxsize=12cm \epsfbox{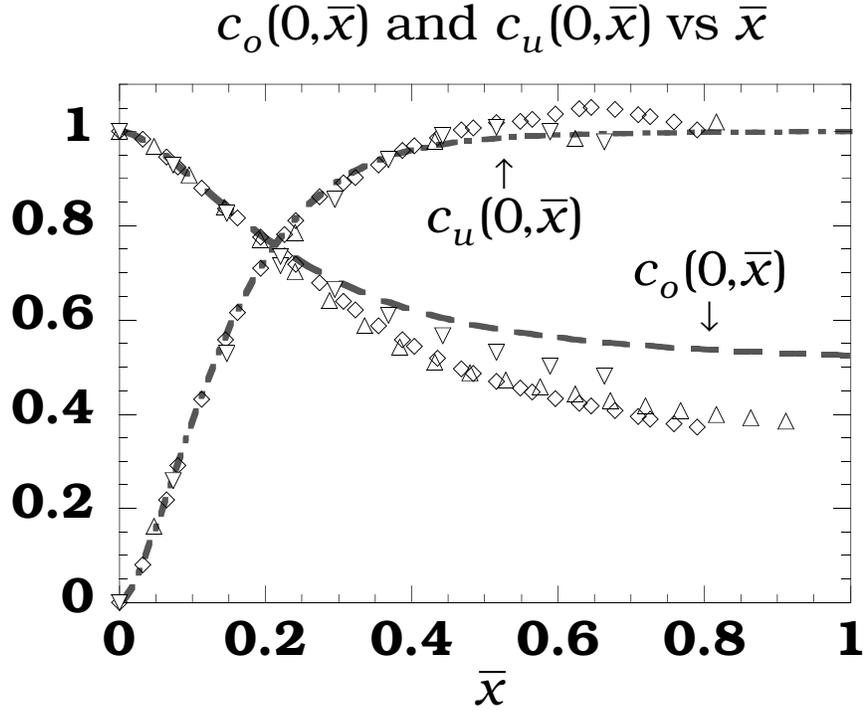}}
\vspace{2cm}
\caption{
  Comparison between numerical and analytical calculations of
  $c_{o,u}(0;\bar x)$.  Numerical calculations of $c_{o,u}$ were obtained
  from $25\times 25(\Diamond)$, $50\times 50(\bigtriangleup)$ and
  $100\times 100(\bigtriangledown)$ random matrices $H(X_o,X_u)=H_0+X_o
  H_s+X_u H_a$ described in the text.  The dot-dashed and the dashed lines
  are analytical results for $c_u(0;\bar x)$ and $c_o(0;\bar x)$ given by
  Eqs.~(\protect\ref{co-result},b).}
\label{fig1}
\end{figure}

In Fig.~\ref{fig1}, we compare our analytical results of $c_{o,u}$ with
numerical results, obtained from the random
matrix $H=H_0+X_oH_s+X_uH_a$, where $H_0$, $H_{s}$ and $H_{a}$ were
defined previously.  
$\left\langle {(\partial E_n/
\partial X_{o,u})^2}\right\rangle$ was determined numerically and rescaled
according to Eqs.~(\ref{def:omega}--\ref{def:C0}).  As is seen in
Fig.~\ref{fig1}, agreement with Eqs.~(\ref{co-result},b) is
good, particularly for small $\bar x$.

So far, we have used the rescaled parameters which are determined in the
unitary or orthogonal limit.  However, in experimental situations, only
the results in the crossover region may be obtained.
For such a case, we propose here the local normalization scheme.  While
$\bar x$ and $x_u$ are formerly defined by Eqs.~(\ref{def:x}), what is
actually observed at a fixed $\bar X$ is
\begin{equation}
  \bar y^2=\left\langle {\left( {{{\partial\varepsilon_n} \over {\partial
          X_u}}} \right)^2} \right\rangle_{\bar X}\bar X^2;\quad\quad
  y_{o,u}^2=\left\langle {\left( {{{\partial\varepsilon_n} \over {\partial
          X_{o,u}}}} \right)^2} \right\rangle_{\bar X}X_{o,u}^2.
\end{equation}
To express all universal properties in terms of $\bar y$ and $y_u$ instead
of $\bar x$ and $x_u$, we can make use of the relation between $\bar y$
and $\bar x$ ($y_{o,u}$ and $x_{o,u}$) and write
\begin{equation}
\bar x^2c_u(\bar x)=\bar y^2;\quad\quad x_{o,u}^2c_{o,u}(\bar x)=y_{o,u}^2.
\end{equation}
By calculating the function $\bar x=f(\bar y)$ and $x_{o,u}=y_{o,u}/
c_{o,u}(f(\bar y))$, we can express the universal dependence in terms of
$\{y_o,y_u,\bar y\}$ rather than $\{x_o,x_u,\bar x\}$, which enables us to
rescale the external parameters through a local point in the crossover
region.

In conclusion, we have presented an analytical expression for the
universal parametric correlations of density of states both in the
orthogonal-unitary and symplectic-unitary crossover region of quantum
chaotic systems. 
For the former, the velocity distribution functions both in the unitary
and the orthogonal direction were presented analytically.  In
particular, the velocity distribution in the unitary direction shows
continuous transition between a Gaussian and a $\delta$-function.  Their
variances both in the orthogonal and unitary directions are shown to have
the logarithmic dependence around the orthogonal limit.  In addition, we
have also proposed a local rescaling scheme of data in the crossover
region.

The authors are grateful to A. Altland, A. Andreev, K. B. Efetov, C.
Itzykson, S. Iida, V. N. Prigodin, and U. Sivan for helpful and
stimulating discussions.  The work was supported by NSF Grant No. DMR
92-04480.  N.T. also acknowledges the research fellowship from Murata
Overseas Scholarship Foundation.

%
%


\begin{references}
\bibitem{supermatrix} 

  Although the parametric extension is straightforward in the supermatrix
  method, the actual complexity of the analytical calculation will highly
  depend on how the saddle-point manifold $Q$ is parametrized.  The
  parametrization used for deriving our main result Eq.(\ref{k-result}) is
  the original one introduced in Ref.~\cite{Efetov83} and slightly
  modified according to suggestions in the appendix of~\cite{Altland92}.
  We remark that parametrization introduced in Ref.~\cite{Altland93} leads
  to a more cumbersome representation of Eq.(\ref{k-result}).
\end{references}

\end{document}